\documentstyle[a4wide,12pt,epsf]{article}
\begin{document}
\begin{flushright}
\baselineskip=12pt
{SUSX-TH-02-021}\\
{hep-th/0210219}\\
{October  2002}
\end{flushright}
\def\IZ{Z\kern-.5em Z}
\begin{center}
{\LARGE \bf STANDARD-LIKE MODELS 
FROM INTERSECTING D5-BRANES \\}
\vglue 0.35cm
{D. BAILIN$^{\clubsuit}$ \footnote
{D.Bailin@sussex.ac.uk}, G. V. KRANIOTIS$^{\spadesuit}$ \footnote
 {G.Kraniotis@sussex.ac.uk, kraniotis@physik.uni-halle.de} and A. LOVE$^{\diamondsuit}$ \\}
\vglue 0.2cm
        {$\clubsuit$ \it  Centre for Theoretical Physics, University of Sussex\\}
{\it Brighton BN1 9QJ, U.K. \\}
{$\spadesuit$ \it Fachbereich Physik, Martin-Luther-Universit\"{a}t Halle-Wittenberg\\
Friedemann-Bach-Platz 6, D-06099, Halle, Germany\\}
{$\diamondsuit$ \it  Centre for Particle Physics, Royal Holloway, University of London \\}
{\it Egham,  Surrey TW20-0EX, U.K. }
\baselineskip=12pt

\vglue 2.5cm
ABSTRACT
\end{center}
We construct intersecting D5-brane orbifold models that yield the (non-supersymmetric) standard model 
up to vector-like matter and charged-singlet scalars.
 The models are constrained by the requirement that twisted tadpoles cancel, 
and that the gauge boson coupled to the weak hypercharge $U(1)_Y$ does not get a string-scale mass via a generalised 
Green-Schwarz mechanism. Gauge coupling constant ratios close to those measured are easily obtained for reasonable 
values of the parameters, consistently with having the string scale close to the electroweak scale,
 as required to avoid the hierarchy problem.

{\rightskip=3pc
\leftskip=3pc
\noindent
\baselineskip=20pt

}

\vfill\eject
\setcounter{page}{1}
\pagestyle{plain}
\baselineskip=14pt

The D-brane world offers an attractive, bottom-up route to getting standard-like models from Type II string theory \cite{UTCA}. 
Open strings that begin and end on a stack of $M$ D-branes generate the gauge bosons of the group $U(M)$ living in the world volume of the D-branes.
 So the 
standard approach is to start with one stack of 3 D-branes, another of 2, and $n$ other stacks each having just 1 D-brane, thereby generating 
the gauge group $U(3) \times U(2) \times U(1)^n$. Fermions in bi-fundamental
 representations of the corresponding gauge groups can arise at the intersections of such stacks \cite{BDL}, but to get $D=4$ {\it chiral} fermions 
  the intersecting branes should sit at a singular point in the space transverse to the branes, an orbifold fixed point, for example. In general,
   such configurations yield a non-supersymmetric spectrum, so to avoid the hierarchy problem the
    string scale associated with such models must be no more than a few TeV. Gravitational interactions occur in the bulk ten-dimensional space, and 
    to ensure that the Planck energy has its observed large value, it is necessary that there are large dimensions transverse to the branes \cite{ADD}. 
The D-branes with which we are concerned wrap the 3-space we inhabit and closed 1-, 2- or 3-cycles of a toroidally 
compactified $T^2, \ T^2 \times T^2$ or $T^2 \times T^2 \times T^2$ space. Thus getting the correct Planck scale effectively means that only D4- and D5-brane models are viable, 
since for D6-branes there is no dimension transverse to all of the intersecting branes.  
In a non-supersymmetric theory the cancellation of the closed-string (twisted) Ramond-Ramond (RR) tadpoles does 
{\it not} ensure the cancellation of the Neveu-Schwarz-Neveu-Schwarz (NSNS) tadpoles.
 There is a resulting instability in the complex structure moduli \cite{BKLO}.
 One  way to stabilise some of the (complex structure) moduli is to use an orbifold, rather than a torus, 
for the space wrapped by the D-branes. If the embedding is supersymmetric, then the instabilities are removed. This has been studied \cite{Cvetic},
 using D6-branes, 
 but it has so far proved difficult to get realistic phenomenology consistent with experimental data from such models.
 Unlike D4-brane models on $T^2 \times T^4/\IZ_N$, D5-brane models on $T^4 \times T^2/\IZ_N$ are necessarily non-supersymmetric 
 in the closed string sector \cite{AFIRU1} and contain closed-string tachyons in the twisted sector. 
 These tachyons may be a source of instability of the background \cite{AFIRU2, CIM}. We have nothing to add to current understanding of this point.

During the past year orientifold models with intersecting D6- and D5-branes
have been constructed that yield 
precisely the fermionic spectrum of the standard model (plus three generations of right-chiral neutrinos) \cite{CIM, IMR}. 
(Other recent work on  intersecting brane models, both supersymmetric 
and non-supersymmetric, and their phenomenological implications may be found in \cite{IBM}.)
The spectrum includes open-string 
 $SU(2)_L$ doublet scalar tachyons that may be regarded as the Higgs doublets that break the electroweak symmetry group, but also, unavoidably,
  open string colour-triplet and charged singlet tachyons either of which is potentially fatal for the phenomenology. 
The wrapping numbers of the various stacks are constrained by the requirement 
 of RR tadpole cancellation, and this ensures the absence of non-abelian anomalies in the emergent low-energy quantum field theory. 
 A generalised Green-Schwarz mechanism ensures that that the gauge bosons 
  associated with all anomalous $U(1)$s acquire string-scale masses \cite{IRU}, but the gauge bosons of some non-anomalous  $U(1)$s
   can also acquire string-scale masses \cite{IMR}; in all such cases the $U(1)$ group survives as a global symmetry.
 Thus we must also   
  ensure the weak hypercharge group $U(1)_Y$ remains a gauge symmetry by requiring that its gauge boson does {\it not} get such a mass. 
 In a recent paper \cite{BKL2} we constructed the first semi-realistic intersecting D4-brane orbifold models that satisfy these constraints.
  We found a unique, one-parameter $(n_2)$ family of six-stack, intersecting D4-brane models having three chiral generations
of matter all of which are coupled to the tachyonic Higgs bosons that generate masses when the electroweak symmetry is spontaneously broken.
Such models all have extra vector-like leptons, as well as open-string charged-singlet scalar tachyons.
 They also have at least one surviving (unwanted) coupled, gauged $U(1)$ symmetry after spontaneous symmetry breaking. 
Relaxing the requirement that all matter has the Higgs couplings necessary to generate masses at renormalisable level
 allows us to construct models without unwanted $U(1)$ gauge symmetries. In this case there are also 
  colour-triplet scalar tachyons as well as vector-like $d$ quark matter.  
Demanding both that there are mass terms at renormalisable level for all matter and that there are no  open-string charged-singlet tachyons 
generally requires models with at least eight stacks of intersecting branes. 
These models predict ratios of the gauge coupling constants that for some values of a parameter are very close to the measured values. 
For these models the string scale must be close to the electroweak scale, and stringy signatures should appear below this scale. 

Intersecting D5-brane models differ in several ways from D4-brane models, 
and it is natural to wonder whether orbifold models with them can do better than with intersecting D4-branes, especially since D5-brane orientifold models 
have been so successful. 
We start with an array of D5-brane stacks, each wrapping  closed 
1-cycles in both 2-tori of $T^2 \times T^2$, and situated
 at a fixed point of the transverse $T^2 /\IZ_3$ orbifold.  
 A stack $a$ is specified by two pairs of wrapping numbers $(n_a,m_a)$  and $(\tilde{n}_a,\tilde{m}_a)$ that 
specify the number of times $a$ wraps the basis 1-cycles  in each  $T^2$. 
When $n_a$ and $m_a$ are coprime 
a single copy of the gauge group $U(N_a)$ occurs;  if $n_a$ and $m_a$ have a highest common factor $f_a$ there are $f_a$ copies of  $U(N_a)$, 
or if $m_a$ (or $n_a$) is zero, then there are $n_a$ (or $m_a$) copies of $U(N_a)$.  The 
same thing happens if  $\tilde{n}_a$ and $\tilde{m}_a$  are not coprime.
The number of intersections of stack $a$ with stack $b$ is given by 
  ${\bf I}_{ab}=I_{ab}\tilde{I}_{ab}$, where $I_{ab} \equiv n_am_b-m_an_b$ and 
  $\tilde{I}_{ab}=\tilde{n}_a\tilde{m}_b-\tilde{m}_a\tilde{n}_b$ are the intersection numbers of the corresponding 1-cycles in each torus $T^2$. 
  The generator $\theta$ of the $Z_3$ point group is embedded in the stack of $N_a$ branes as
 $\gamma _{\theta,a} = \alpha ^{p_a}I_{N_a}$, where $\alpha = e^{2\pi i/3}, \ p_a=0,1,2$. 
 The first two stacks $a=1,2$ defined above, that generate a $U(3) \times U(2)$ gauge group, are common to all models. 
 We require that the intersection  
 of these two stacks gives the three copies of the left-chiral quark doublet $Q_L$. Then, without loss of generality, 
 we may take their wrapping numbers as shown in Table 1. 
 We require that $n_2$ is coprime to 3 ($n_2 \neq 0 \bmod 3$), so that the $U(2)$ group is not replicated.
  Besides the first two stacks we have, in general,  
 three sets $I,J,K$ of $U(1)$ stacks characterised by their Chan-Paton factors: $p_i=p_2  \ \forall i \in I, \ p_1 \neq p_j \neq p_2  \ \forall j \in J$, and 
 $p_k =p_1  \ \forall k \in K$; the sets $I,J,K$ are each divided into two subsets $I_1 \cup I_2=I$ etc.,
  defined \cite{BKL1} so that the weak hypercharge $Y$ is the linear 
 combination 
 \begin{equation}
 -Y=\frac{1}{3}Q_1+\frac{1}{2}Q_2+ \sum_{i_1 \in I_1}Q_{i_1}+ \sum_{j_1 \in J_1}Q_{j_1}+ \sum_{k_1 \in K_1}Q_{k_1}
 \label{Y}
 \end{equation}
where $Q_a$ is the $U(1)$ charge associated with the stack $a$; $Q_a$ is normalised such that the ${\bf N}_a$ 
 representation of $SU(N_a)$ has $Q_a=+1$. 
  \begin{table}
\begin{center}
\begin{tabular}{|c|c|c|c|c|} \hline \hline
Stack $a$ & $N_a$ & $(n_a,m_a)$ & $(\tilde{n}_a,\tilde{m}_a)$ & $\gamma_{\theta,a}$ \\
\hline \hline
1 & 3 & $(1,0)$ & $(1,0)$ & $\alpha ^p {\bf I}_3$ \\
2 & 2 & $(n_2,3)$ & $(\tilde{n}_2,1)$ & $\alpha ^q {\bf I}_2 $\\
$i \in I$ & 1 & $(n_{i},m_{i})$ & $(\tilde{n}_i,\tilde{m}_i)$ & $\alpha ^q $\\
$j\in J$ & 1 & $(n_{j},m_{j})$ & $(\tilde{n}_j,\tilde{m}_j)$ & $\alpha ^r $\\
$k \in K$ & 1 & $(n_{k},m_{k})$ & $(\tilde{n}_k,\tilde{m}_k)$ & $\alpha ^p$ \\
\hline \hline
\end{tabular}
\end{center}
\caption{Multiplicities, wrapping numbers and Chan-Paton phases for the D5-brane models, 
($p\neq q \neq r \neq p$).}
\end{table}

The cancellation of twisted tadpoles gives the constraints:
\begin{eqnarray}
\sum_a n_a\tilde{n}_a{\rm trace}\gamma _{\theta,a}=0   \label{nnt} \\
\sum_a n_a\tilde{m}_a{\rm trace}\gamma _{\theta,a}=0   \label{nmt} \\
\sum_a m_a\tilde{n}_a{\rm trace}\gamma _{\theta,a}=0   \label{mnt}  \\
\sum_a m_a\tilde{m}_a{\rm trace}\gamma _{\theta,a}=0   \label{mmt}
\end{eqnarray}
and this is sufficient to ensure cancellation of the cubic non-abelian anomalies. For the D5-brane array given in Table 1, 
the first of these constraints (\ref{nnt}) gives
\begin{equation}
2n_2\tilde{n}_2+\sum_{i \in I}n_i\tilde{n}_i=\sum_{j \in J}n_j\tilde{n}_j=3+\sum_{k \in K}n_k\tilde{n}_k
\label{nn}
\end{equation}
 The mixed $U(1)$ anomalies are cancelled by
 a generalised Green-Schwarz mechanism involving the exchange of fields that arise from the dimensional reduction of the twisted two-form RR fields 
 $B_2^{(k)}, \ C_2^{(k)}, \ D_2^{(k)}$ and $E_2^{(k)}$ that 
 live at the orbifold singularity \cite{AFIRU1}. These fields are coupled to the $U(1)_a$ field strength $F_a$ of the stack $a$ by terms in the low energy 
 action of the form 
 \begin{eqnarray}
 n_a \tilde{n}_a \int _{M_4}{\rm Tr}(\gamma _{k,a} \lambda _a)B_2^{(k)} \wedge {\rm Tr}F_a  \label{nnB}\\
 m_a \tilde{m}_a\int _{M_4}{\rm Tr}(\gamma _{k,a} \lambda _a)C_2^{(k)} \wedge {\rm Tr}F_a  \\
 m_a \tilde{n}_a \int _{M_4}{\rm Tr}(\gamma _{k,a} \lambda _a)D_2^{(k)} \wedge {\rm Tr}F_a \\
 n_a \tilde{m}_a \int _{M_4}{\rm Tr}(\gamma _{k,a} \lambda _a)E_2^{(k)} \wedge {\rm Tr}F_a 
 \end{eqnarray}
 where $\gamma _{k,a} \equiv \gamma _{\theta,a}^k $ and $\lambda _a$ is the Chan-Paton matrix associated with the $U(1)$ generator. 
 The couplings determine the linear combination of $U(1)$ gauge bosons that acquire string-scale masses via the Green-Schwarz mechanism. 
 For the D5-brane array given in Table 1 the coupling (\ref{nnB}) to $B_2^{(k)}$ is
 \begin{eqnarray}
 \left[ \alpha^{pk} \left( 3F_1+\sum_{k_1}n_{k_1}\tilde{n}_{k_1}F_{k_1} +\sum_{k_2}n_{k_2}\tilde{n}_{k_2}F_{k_2}-\sum_{j_1}n_{j_1}\tilde{n}_{j_1}F_{j_1}
 -\sum_{j_2}n_{j_2}\tilde{n}_{j_2}F_{j_2} \right)  \right. \nonumber \\
 \left. + \alpha^{qk} \left( 2n_2\tilde{n}_{2}F_2+\sum_{i_1}n_{i_1}\tilde{n}_{i_1}F_{i_1} +\sum_{i_2}n_{i_2}\tilde{n}_{i_2}F_{i_2}
 -\sum_{j_1}n_{j_1}\tilde{n}_{j_1}F_{j_1}-\sum_{j_2}n_{j_2}\tilde{n}_{j_2}F_{j_2} \right) \right] \wedge B_2^{(k)} 
 \label{B2k}
  \end {eqnarray}
We require that the $U(1)_Y$ gauge boson associated with the weak hypercharge given in eqn (\ref{Y}) remains massless. Consequently, the corresponding 
field strength must be orthogonal to those that acquire Green-Schwarz masses. Thus we require that the wrapping numbers satisfy the constraints:
\begin{equation}
n_2\tilde{n}_2+ \sum _{i_1}n_{i_1}\tilde{n}_{i_1}   =   \sum _{j_1}n_{j_1}\tilde{n}_{j_1} = 1+ \sum _{k_1}n_{k_1}\tilde{n}_{k_1} \equiv T_{11} \label{T11}  
\end{equation}
Combining these with the tadpole cancellation constraints (\ref{nn}) gives also
 \begin{eqnarray}
n_2\tilde{n}_2+ \sum _{i_2}n_{i_2}\tilde{n}_{i_2}   =  \sum _{j_2}n_{j_2}\tilde{n}_{j_2} = 2+ \sum _{k_2}n_{k_2}\tilde{n}_{k_2} \equiv T_{12} \label{T12} 
\end{eqnarray}
It follows that the sets $J_1 \cup K_1$ and $J_2 \cup K_2$ are both non-empty.
Requiring also that the weak hypercharge field strength is othogonal to the other combinations that acquire Green-Schwarz masses, we obtain the 
constraints
\begin{eqnarray}
3+ \sum _{i_1}m_{i_1}\tilde{m}_{i_1}   =   \sum _{j_1}m_{j_1}\tilde{m}_{j_1} =  \sum _{k_1}m_{k_1}\tilde{m}_{k_1} \equiv T_{41}   \label{T41}  \\
3+ \sum _{i_2}m_{i_2}\tilde{m}_{i_2}   =   \sum _{j_2}m_{j_2}\tilde{m}_{j_2} =  \sum _{k_2}m_{k_2}\tilde{m}_{k_2} \equiv T_{42}  \label{T42} \\
n_2 +\sum _{i_1}n_{i_1}\tilde{m}_{i_1}   =   \sum _{j_1}n_{j_1}\tilde{m}_{j_1} =  \sum _{k_1}n_{k_1}\tilde{m}_{k_1} \equiv T_{21}   \label{T21}  \\
n_2+ \sum _{i_2}n_{i_2}\tilde{m}_{i_2}   =   \sum _{j_2}n_{j_2}\tilde{m}_{j_2} =  \sum _{k_2}n_{k_2}\tilde{m}_{k_2} \equiv T_{22} \label{T22} \\
3\tilde{n}_2+ \sum _{i_1}m_{i_1}\tilde{n}_{i_1}   =   \sum _{j_1}m_{j_1}\tilde{n}_{j_1} =  \sum _{k_1}m_{k_1}\tilde{n}_{k_1} \equiv T_{31}   \label{T31} \\   
3\tilde{n}_2+ \sum _{i_2}m_{i_2}\tilde{n}_{i_2}   =   \sum _{j_2}m_{j_2}\tilde{n}_{j_2} =  \sum _{k_2}m_{k_2}\tilde{n}_{k_2} \equiv T_{32}  \label{T32} 
\end{eqnarray}

Tachyonic scalars arise only at intersections between stacks $a$ and $b$ which have the same Chan-Paton factor $p_a=p_b$. Thus,
 Higgs doublets, which are needed to give mass to the fermionic matter, arise at $(2i_1)$ and $(2i_2)$ intersections.  
 With D5-branes, chiral fermions arise only at intersections with $p_a \neq p_b$. Thus to obtain (doublet and singlet) leptons, we also require  
 $U(1)$ stacks in the sets $J$ and/or $K$. We consider here models with not more than 
 one stack in each of the classes $I_1,I_2,J_1,J_2,K_1,K_2$. Then the consistency of eqns
  (\ref{T11},\ref{T41},\ref{T21},\ref{T31}) 
 requires that 
 \begin{eqnarray}
 (T_{21}-n_2)(T_{31}-3\tilde{n}_2)=n_{i_1}\tilde{m}_{i_1}m_{i_1}\tilde{n}_{i_1}=(T_{11}-n_2\tilde{n}_2)(T_{41}-3) \\
 T_{21}T_{31}=n_{j_1}\tilde{m}_{j_1}m_{j_1}\tilde{n}_{j_1}=T_{11}T_{41} \\
  T_{21}T_{31}=n_{k_1}\tilde{m}_{k_1}m_{k_1}\tilde{n}_{k_1}=(T_{11}-1)T_{41}
 \end{eqnarray}
 so that 
 \begin{equation}
 T_{41}=0=T_{21}T_{31}=3T_{11}-n_2T_{31}-3\tilde{n}_2T_{21}
 \label{T410}
 \end{equation}
 Using the first of these the $(2i_1)$ intersection numbers are 
 \begin{eqnarray}
 I_{2i_1} \equiv n_2m_{i_1}-3n_{i_1}=T_{21}m_{i_1} \\
 \tilde{I}_{2i_1} \equiv \tilde{n}_2\tilde{m}_{i_1}-\tilde{n}_{i_1}=\frac{1}{3}T_{31}\tilde{m}_{i_1}
 \end{eqnarray}
 so using the second of eqns (\ref{T410}) we see that at least one of the intersection numbers is zero. 
 In the same way we can show that 
 \begin{equation}
 T_{42}=0=T_{22}T_{32}=3T_{12}-n_2T_{32}-3\tilde{n}_2T_{22}
 \label{T420}
 \end{equation}
 and hence that ${\bf I}_{2i_2} \equiv I_{2i_2}\tilde{I}_{2i_2}=0$. However, unlike the situation for intersecting D4-brane models, 
 this does not necessarily imply that such models have no 
 Higgs doublets. The vanishing of the intersection number on one of the tori means that on that torus the 1-cycles wrapped by the two stacks 
 are parallel. Provided that the 1-cycles on the other torus do intersect, scalar doublets arise, and for sufficiently small separation between 
 the parallel 1-cycles, these states are tachyonic \cite{IMR,CIM}. Thus, in general in D5-brane models,  
we obtain scalar 
 tachyons at intersections of stacks $a$ and $b$ (with $p_a=p_b$) for which $I_{ab}$ and $\tilde{I}_{ab}$ are  both non-zero, and we have the freedom 
  to obtain scalar 
 tachyons at intersections where either $I_{ab}$ or $\tilde{I}_{ab}$, but not both, are zero.

 To ensure that Higgs doublets arise at the $(2i_1)$ intersection, either $T_{21}$ or $T_{31}$, must be non-zero.
 Suppose first that $T_{21}=0$, so from (\ref{T410}) $n_2T_{31}=3T_{11} \neq 0$. Since $n_2 \neq 0 \bmod 3$, it follows that $T_{31}=3p$
  with $p \neq 0$ an integer, and $T_{11}=n_2p$. Then eqn (\ref{T11}) shows that $n_{j_1} \neq 0$, and eqn (\ref{T21})
   requires $\tilde{m}_{j_1}=0$. We shall see shortly that the replication of any stack gives too many $U(1)$ factors in the gauge group 
   to yield just the standard model.    
   To avoid replication of the $j_1$ stack we therefore require $\tilde{n}_{j_1} = \epsilon _{j_1} \equiv \pm 1$. 
   We can then solve eqns (\ref{T11}, \ref{T32}) to give $n_{j_1}=\epsilon _{j_1}n_2p$ and $m_{j_1}=\epsilon _{j_1}3p$. To avoid replication we 
   therefore require that the common factor $p= \pm 1$. The arbitrary overall sign factor $\epsilon _{j_1}$ in each of the four wrapping numbers
    of the stack represents a symmetry which is present for all stacks, since changing the sign of $n_a,m_a,\tilde{n}_a$ and $\tilde{m}_a$
     for any of the stacks $a \in I \cup J \cup K$ leaves eqns (\ref{T11}) .. (\ref{T32}) invariant. The total intersection number
     ${\bf I}_{ab}$ is unaffected by these phases, and so therefore is the spectrum. Proceeding in this way, 
     we may similarly solve for the wrapping numbers of the $j_1$ and $k_1$ stacks. The solution is given in the top half of Table 2. The bottom 
     half gives the solution for these stacks when $T_{31}=0, \ T_{21}=p \neq 0$ and $T_{11}=\tilde{n}_2p$. To avoid triplication of gauge group facors, 
     we require in either case that 
     \begin{equation}
     n_2=-p \bmod 3
     \label{p}
     \end{equation}      
     Table 3 gives the analogous results for the $i_2,j_2$ and $k_2$ stacks, and to avoid triplication we require
     \begin{eqnarray}
     n_2=q \bmod 3 \ {\rm for} \ T_{22}=0  \label{q1}\\
     n_2=-q \bmod 3 \ {\rm for} \ T_{32}=0  \label{q2}
     \end{eqnarray}
  \begin{table}
\begin{center}
\begin{tabular}{|c|c|c|c|c|} \hline \hline
Stack $a$ &  $n_a$ & $m_a$ & $\tilde{n}_a$ & $\tilde{m}_a$  \\
\hline \hline
$i_1 $ & $n_2$ & $3$ & $p-\tilde{n}_2$ & $-1$ \\
$j_1 $ & $n_2p$ & $ 3p$ & $1 $ & $0$  \\
$k_1 $  & $n_2p-1$ & $3p$ & $1$ & $0$ \\
\hline 
$i_1$ & $n_2-p$ & $3$ & $-\tilde{n}_2$ & $-1$ \\
$j_1$ &  $1$ & $ 0$ & $\tilde{n}_2p$ & $p$ \\
$k_1$ &   $1$ & $0$ & $\tilde{n}_2p-1$ & $p$ \\
\hline \hline
\end{tabular}
\end{center}
\caption{Wrapping numbers for the stacks $i_1,j_1$ and $k_1$. At the top $T_{21}=0, \ T_{31}=3p, \ T_{11}=n_2p$, 
and at the bottom $T_{21}=p, \ T_{31}=0, \ T_{11}=\tilde{n}_2p$. In both cases $p= \pm 1$. A further overall, arbirary 
sign $\epsilon _a = \pm 1$ is understood for each stack $a=i_1,j_1,k_1$.}
\end{table}
  
  \begin{table}
\begin{center}
\begin{tabular}{|c|c|c|c|c|} \hline \hline
Stack $a$ &  $n_a$ & $m_a$ & $\tilde{n}_a$ & $\tilde{m}_a$  \\
\hline \hline
$i_2 $ & $n_2$ & $3$ & $q-\tilde{n}_2$ & $-1$ \\
$j_2 $ & $n_2q$ & $3q$ & $1$ & $0$  \\
$k_2 $  & $n_2q-2$ & $3q$ & $1$ & $0$ \\
\hline 
$i_2$ & $q-n_2$ & $-3$ & $\tilde{n}_2$ & $1$ \\
$j_2$ &  $1$ & $0$ & $\tilde{n}_2q $ & $q$ \\
$k_2$ &   $1$ & $0$ & $\tilde{n}_2q-2$ & $q$ \\
\hline \hline
\end{tabular}
\end{center}
\caption{Wrapping numbers for the stacks $i_2,j_2$ and $k_2$. At the top $T_{22}=0, \ T_{32}=3q, \ T_{12}=n_2q$, 
and at the bottom $T_{22}=q, \ T_{32}=0, \ T_{12} =\tilde{n}_2 q$. In both cases $q= \pm 1$. A further arbitrary, overall 
sign $\epsilon _a = \pm 1$ is understood for each stack $a=i_2,j_2,k_2$.}
\end{table}

 Evidently there are four classes of models 
obtained by combining one of the options in Table 2 with one of the options in Table 3.
For no choice of the parameters can any  of the stacks in Table 2 or 3 be absent.
Thus a minimum of eight stacks is required, and in fact this a general result, 
even if we allow more than one stack in some of the sets \cite{BKL4}.
In the first instance all such eight-stack models have gauge group $U(3) \times U(2) \times U(1)^6$. 
By construction the standard model gauge group $SU(3)_c \times SU(2)_L 
\times U(1)_Y$ survives the Green-Schwarz mechanism that gives string-scale masses to some of the original (eight) $U(1)$ gauge bosons. In addition, 
it is easy to see that the symmetry $U(1)_X$, associated with the the sum of the charges 
\begin{equation}
X=\sum_a Q_a
\end{equation}
also survives as a gauge symmetry. However, this is uncoupled to all of the matter and gauge 
fields, and so is physically unobservable. It remains to determine whether any other $U(1)$s survive as gauge symmetries. Couplings to the 
form  $B_2^{(k)}$ in (\ref{B2k}) give Green-Schwarz masses to the gauge bosons of the $U(1)$ symmetries 
associated with the two linear combinations of $U(1)$ charge:
\begin{eqnarray}
3Q_1-Q_{k_1}-2Q_{k_2} +T_{11}(Q_{k_1}-Q_{j_1}) +T_{12}(Q_{k_2}-Q_{j_2}) \\
n_2\tilde{n}_2(2Q_2-Q_{i_1}-Q_{i_2}) +T_{11}(Q_{i_1}-Q_{j_1}) +T_{12}(Q_{i_2}-Q_{j_2}) \label{B2}
\end{eqnarray}
Using eqns (\ref{T410},\ref{T420}), the couplings to  $C_2^{(k)}$ give a Green-Schwarz mass 
only to the gauge boson of the single $U(1)$ symmetry associated with the linear combination:
\begin{equation}
2Q_2-Q_{i_1}-Q_{i_2}
\label{C}
\end{equation}
For general values of $T_{21},T_{22},T_{31}$ and $T_{32}$ the remaining couplings to $D_2^{(k)}$ and $E_2^{(k)}$ 
generate Green-Schwarz masses for the combinations:
\begin{eqnarray}
n_2(2Q_2-Q_{i_1}-Q_{i_2}) + T_{21}(Q_{i_1}-Q_{j_1})+ T_{22}(Q_{i_2}-Q_{j_2}) \label{D1}\\ 
T_{21}(Q_{k_1}-Q_{j_1})+ T_{22}(Q_{k_2}-Q_{j_2})  \label{D2} \\
3\tilde{n}_2(2Q_2-Q_{i_1}-Q_{i_2}) + T_{31}(Q_{i_1}-Q_{j_1})+ T_{32}(Q_{i_2}-Q_{j_2}) \label{E1} \\
T_{31}(Q_{k_1}-Q_{j_1})+ T_{32}(Q_{k_2}-Q_{j_2}) \label{E2}
\end{eqnarray} 

Consider first the models obtained when $T_{21}=0=T_{22}$, corresponding to choosing the top option in both tables. Then the combination (\ref{D2}) 
vanishes, (\ref{D1}) reduces to (\ref{C}), and (\ref{E1}) reduces to  (\ref{B2}). Thus only 4 of the original 8 $U(1)$ gauge bosons get Green-Schwarz 
masses, leaving 4 massless, gauged $U(1)$s. The same is true of the models with  $T_{31}=0=T_{32}$, corresponding to choosing the bottom option in 
both tables. We know that two of the surviving $U(1)$s are $U(1)_X$ and $U(1)_Y$. The latter is spontaneously broken when the Higgs doublets 
at the $(2i_1)$ and $(2i_2)$ intersections acquire non-zero VEVs. Clearly not more than one further $U(1)$ symmetry can be broken by these VEVs, so 
there remains at least one, unwanted gauged $U(1)$ symmetry that survives the electroweak symmetry breaking, and which is coupled to the standard 
model matter.  

If instead we choose $T_{21}=0=T_{32}$, then in this case 6  of the original 8 $U(1)$ gauge bosons get Green-Schwarz 
masses, leaving just the massless, gauged $U(1)_X$ and $U(1)_Y$. Potentially this can  yield  a standard-like model. 
The consistency of the conditions (\ref{p}) and (\ref{q2}) for avoiding triplication requires
\begin{equation}
n_2=-p \bmod 3=-q \bmod 3
\end{equation}
It is straightforward to determine the spectrum, which is independent of $p$. 
As before in the D4-brane case \cite{BKL1}, the  three generations of chiral matter include right-chiral neutrino states, and
 there is additional vector-like leptonic, but not quark, matter.
 We find
\begin{equation}
3(L+\bar{L})+6(e^c_L +\bar{e}^c_L)+3(\nu^c_L +\bar{\nu}^c_L)
\label{ng31}
\end{equation}
 There is also one Higgs doublet at the $(2i_1)$ intersections, and 3 Higgs doublets at the $(2i_2)$ intersections. Less welcome
  are the 9 charged-singlet tachyons that arise, 3 at each of the $(i_1i_2), \ (j_1j_2)$ and $(k_1k_2)$ intersections. 
  These all arise from intersections where both intersection numbers $ I_{ab}$ and  $\tilde{I}_{ab}$ are non-zero,
   and so, as noted earlier, the tachyons cannot be removed from the spectrum by increasing the separation of parallel 1-cycles.
   In general, the squared mass of a tachyon at an intersection of stack $a$ with stack $b$ is given by \cite{AFIRU1}
   \begin{equation}
   m_{ab}^2=-\frac{m_{\rm string}^2}{2\pi}\left| \frac{\epsilon I_{ab}R_2/R_1}{|n_a - m_aR_2/R_1||n_b - m_bR_2/R_1|}-
   \frac{\tilde{\epsilon} \tilde{I}_{ab}\tilde{R}_2/\tilde{R}_1}{|\tilde{n}_a - \tilde{m}_a\tilde{R}_2/\tilde{R}_1||\tilde{n}_b - 
   \tilde{m}_b\tilde{R}_2/\tilde{R}_1|}\right|
   \end{equation}
where $R_1,R_2,\tilde{R}_1$ and $\tilde{R}_2$ are the radii of the fundamental 1-cycles on the two tori on which the D5-branes are wrapped;
\begin{equation}
 \epsilon \equiv 2|\cos (\theta /2)| \quad {\rm and} \quad  \tilde{\epsilon}\equiv 2|\cos (\tilde{\theta} /2)|
 \end{equation}
 with $\theta$ and $\tilde{\theta}$ the angles between the vectors defining the lattices on the two tilted tori.
 The above formula is valid provided that 
 $\epsilon$ and $\tilde{\epsilon}$ are small;  this is required in any case to ensure that the masses of the Higgs doublets are small 
 compared with the string scale. In principle, the contributions from the two tori can cancel leaving massless states rather than tachyons. 
 For this to happen for the charged-singlet tachyons at the  $(i_1i_2), \ (j_1j_2)$ and $(k_1k_2)$ intersections, the parameters must satisfy 
 the conditions
 \begin{equation}
 \frac{\epsilon R_2/R_1}{\tilde{\epsilon}\tilde{R}_2/\tilde{R}_1}=\left| \frac{x(p-x)}{y(p-y)} \right|=\left| \frac{x}{y} \right| = \left| \frac{p-x}{2p-y} \right|
 \end{equation}
 where 
\begin{equation}
x \equiv n_2 -3R_2/R_1, \qquad y \equiv \tilde{n}_2-\tilde{R}_2/\tilde{R}_1
\end{equation}
 These are all satisfied if 
 \begin{equation}
 px=\frac{3}{2}=py \quad {\rm or \ if } \quad px=\frac{2}{3}=\frac{1}{2}py
 \label{mc1}
 \end{equation}
  The 3 colour-triplet scalars that arise at the $(1k_1)$ intersection, as well as the one at the $(1k_2)$ intersection, can be expunged by 
   taking the separation between the parallel 1-cycles to be sufficiently large. 
    The allowed Yukawas satisfy selection rules that derive from a $Z_2$ symmetry associated with each stack of D5-branes.
 A state associated with a string 
between the $a$th and $b$th stack of D5-branes is odd under 
the $a$th and $b$th $Z_2$ and even under any other $Z_2$. It is easy to see that this selection rule allows the Yukawa couplings 
of the Higgs doublets needed to generate mass terms at renormalisable level for the three generations of chiral matter, including neutrinos. 
However, there are no tree-level Yukawa couplings of the $\bar{e}_L^c$  states.
 None of the vector-like $e_L^c +\bar{e}_L^c$ matter 
acquires a mass at renormalisable level, although the vector-like $\nu _L^c +\bar{\nu}^c_L$ matter does. 
 
 For the case $T_{22}=0=T_{31}$, again only $U(1)_X$ and $U(1)_Y$ evade acquiring Green-Schwarz 
masses and so survive as gauged symmetries. The consistency of (\ref{p}) and (\ref{q1}) now requires that
 \begin{equation}
n_2=-p \bmod 3=q \bmod 3
\end{equation}
In this case we find the vector-like matter to be
\begin{equation}
3(L+\bar{L})+12(e^c_L +\bar{e}^c_L)+3(\nu^c_L +\bar{\nu}^c_L)
\label{ng32}
\end{equation}
The total tachyonic Higgs, colour-triplet and charged-singlet scalar content is the same as for the above model. In this case though,
there are 3  Higgs doublets at the $(2i_1)$ intersections, and one at $(2i_2)$; the colour-triplet scalars are similarly interchanged.
They are massless if 
\begin{equation}
 px=-\frac{3}{2}=-py \quad {\rm or  \ if } \quad px=\frac{4}{3}=-2py
 \label{mc2}
 \end{equation}
As for the previous case, there are Yukawa couplings 
of the Higgs doublets needed to generate mass terms at renormalisable level for the three generations of chiral matter, including neutrinos. 
Some, but not all, of the vector-like $e_L^c +\bar{e}_L^c$ matter 
acquires a mass at renormalisable level, and again all of the vector-like $\nu^c_L +\bar{\nu}^c_L$ matter does.

 Ratios of the gauge coupling constants are independent of the Type II string coupling constant $\lambda_{II}$. Thus
\begin{equation}
\frac{\alpha_3(m_{\rm string})}{\alpha_2(m_{\rm string})}=|xy|
\label{32}
\end{equation}
Also, since 
\begin{equation}
\frac{1}{\alpha_Y} = \frac{1}{3\alpha_3}+\frac{1}{2\alpha_2}+\frac{1}{\alpha_{i_1}}+\frac{1}{\alpha_{j_1}}+\frac{1}{\alpha_{k_1}}
\end{equation}
we have for the $T_{21}=0=T_{32}$ model that
\begin{equation} 
\frac{\alpha_3(m_{\rm string})}{\alpha_Y(m_{\rm string})}=\frac{1}{3}+\frac{1}{2}|xy|+|x(p-y)|+ |x| + |px-1|
\label{3Y}
\end{equation}
Consistency with a low string scale requires that these ratios do not differ greatly from the values measured \cite{cernyellow} at the 
electroweak scale $m_Z$
\begin{eqnarray}
\frac{\alpha_3(m_Z)}{\alpha_2(m_Z)} = 3.54 \\
\frac{\alpha_3(m_Z)}{\alpha_Y(m_Z)} = 11.8 \label{3Yexp}
\end{eqnarray}
It is easy to find all solutions of these:
\begin{eqnarray}
px=4.78, \ -1.75, \ 2.41, \ -4.11 \nonumber \\
py=0.74, \ -2.02, \ -1.46, \ 0.86
\label{pxy}
\end{eqnarray}
which can be satisfied with reasonable values of the parameters. 
For example, $n_2=-2 =1 \bmod 3, \ R_2/R_1=0.93, \   \tilde{n}_2=1, \ \tilde{R}_2/\tilde{R}_1=1.74$.
 None of these solutions is close to satisfying the conditions (\ref{mc1}) for massless charged-singlet scalars, so the existence 
 of these tachyons is unavoidable in this model unless there are very large radiative corrections.

For the $T_{31}=0=T_{22}$ model, eqn (\ref{3Y}) is replaced by
 \begin{equation} 
\frac{\alpha_3(m_{\rm string})}{\alpha_Y(m_{\rm string})}=\frac{1}{3}+\frac{1}{2}|xy|+|(x-p)y|+ |y| + |py-1|
\label{3Y1}
\end{equation}
and the solutions are obtained by interchanging $px$ and $py$ in (\ref{pxy}). Again, it is easy to satisfy these 
with reasonable values of the parameters. For example, $n_2=1, \ R_2/R_1=1.00, \   \tilde{n}_2=-1, \ \tilde{R}_2/\tilde{R}_1=0.75$. 
In this case, the third solution is not too far removed from satisfying the first condition in (\ref{mc2}) and it is 
possible that the effects of renormalisation group running and/or radiative corrections to the masses could remove the charged-singlet 
scalar tachyons from the low-energy spectrum.

In conclusion, we find that requiring that the gauge boson  associated with weak hypercharge does  not acquire 
a string-scale mass requires intersecting D5-brane models with at least eight stacks, if we are to get the standard model gauge group 
with {\em no} additional, unwanted gauged $U(1)$ symmetries, plus three (non-supersymmetric) generations of chiral matter. This parallels 
the situation in orientifold models, in which a minimum of four stacks, plus their orientifold images, is required \cite{CIM}.
In this paper we have studied models with just one 
stack in each of the sets  $I_1,I_2,J_1,J_2,K_1,K_2$, and we found two distinct classes of model.
Both classes of model have extra vector-like leptons, as well as charged-singlet scalar tachyons, and optional colour-triplet tachyons. There 
are radiative corrections to the masses of both that might be sufficient to render them non-tachyonic, without removing the Higgs doublets. 
For  particular values of the wrapping numbers, both classes of model predict ratios of gauge coupling constants 
close to those measured at the electroweak scale, thereby allowing the possibility of  a  nearby string scale, 
such as is required of non-supersymmetric theories to avoid the hierarchy problem. For the model with $T_{31}=0=T_{22}$, one choice of 
the parameters comes close to removing the charged-singlet scalar tachyons from the low-energy spectrum.
Both classes of model possess the Yukawa couplings of the tachyonic Higgs doublets needed to generate masses at renormalisable level for 
three generations of chiral matter including neutrinos. However, neither class of model has masses at renormalisable level for 
all of the vector-like matter. The reason for this undesirable feature is the mismatch in eqns (\ref{ng31}) and (\ref{ng32}) between the number of 
 $L+\bar{L}$ pairs and the number of $e_L^c + \bar{e}_L^c$ pairs. 
 All models have anomalous $U(1)$s that survive as global symmetries. Their gauge bosons acquire string-scale masses via the 
 generalised Green-Schwarz mechanism. It is expected that TeV-scale $Z'$ vector bosons will be observable at future colliders,
  and precision electroweak data (on the $\rho$-parameter) 
  already constrain \cite{GIIQ} the string scale  to be at least 1.5 TeV. 
  In particular, baryon 
  number $B=Q_1/3$ is anomalous and survives as a global symmetry. Consequently, the proton is stable despite the low string scale. 
  As before in the D4-brane case \cite{BKL1},
  the Higgs boson fields are also charged under some of the anomalous $U(1)$s that survive as global symmetries. Thus a keV-scale axion is unavoidable.

 \section*{Acknowledgements}
This research is supported in part by PPARC and the German-Israeli 
Foundation for Scientific Research (GIF).


\newpage

\begin{thebibliography}{99}

\bibitem{UTCA} N. D. Lambert and P. C. West, JHEP 9909 (1999) 021, hep-th/9909129, 
 New J. Phys. 4 (2002), hep-th/0012121;
G. Aldazabal, L. E. Ib\'{a}\~{n}ez and F. Quevedo, JHEP 0001(2000)031, hep-th/9909172,
JHEP 0002 (2000) 015, hep-th/0005067; 
I. Antoniadis, E. Kiritsis and T. Tomaras, Phys. Lett. B486 (2000) 186, hep-th/0004214;
 D. Bailin, G.V. Kraniotis, A. Love, Phys. Lett. B502 (2001) 209, hep-th/0011289;
C. Bachas, hep-th/9503030;   R. Blumenhagen, L. G\"{o}rlich, B. K\"{o}rs and 
D. L\"{u}st, JHEP 0010 (2000) 006, hep-th/0007024;
Z. Kakushadze, Phys. Rev. D59 (1999) 045007;
 Z. Kakushadze, G. Shiu and S.-H. Henry Tye, Nucl. Phys. B533(1998)25;
J. Lykken, E. Poppitz, S. P. Trivedi, Nucl. Phys. B543 (1999) 105; 
I. Antoniadis, E. Dudas, A. Sagnotti, Phys. Lett.B 464 (1999) 38; S. Sugimoto, 
Prog. Theor. Phys. 102 (1999) 685; C. Angelantonj, Nucl. Phys.B 566 (2000) 126;

\bibitem{BDL}M. Berkooz, M. R. Douglas and R. G. Leigh, Nuclear Physics B480 (1996) 265, hep-th/9606139

\bibitem{ADD}N. Arkani-Hamed, S. Dimopoulos and G.R. Dvali, Phys. Lett. B429 (1998) 263, hep-ph/9803315; 
I. Antoniadis, N. Arkani-Hamed, S. Dimopoulos and G.R.Dvali, Phys. Lett.B 436 (1998) 257, hep-ph/9804398.

\bibitem{BKLO}R. Blumenhagen, B. K\"{o}rs, D. L\"{u}st and T. Ott, Nucl. Phys.B616 (2001) 3, hep-th/0107138 

\bibitem{Cvetic}M. Cveti\v{c}, G.Shiu and A. M. Uranga, Nucl. Phys. B615 (2001) 3, hep-th/0107166;
M. Cveti\v{c}, P. Langacker and G. Shiu, hep-ph/0205252, hep-th/0206115

\bibitem{AFIRU1} G. Aldazabal, S. Franco, L. E. Ib\'{a}\~{n}ez, R. Rabad\'{a}n and A. M. Uranga, J. Math. Phys. 42 (2001) 3103, hep-th/0011073

\bibitem{AFIRU2} G. Aldazabal, S. Franco, L. E. Ib\'{a}\~{n}ez, R. Rabad\'{a}n and A. M. Uranga, JHEP 0102 (2001) 047, hep-ph/0011132

\bibitem{CIM}D. Cremades, L. E. Ib\'{a}\~{n}ez and F. Marchesano, hep-th/0205074

\bibitem{IMR}L. E. Ib\'{a}\~{n}ez, F. Marchesano and R. Rabadan, JHEP 0111 (2001) 002, hep-th/0105155

\bibitem{IBM}R. Blumenhagen, V. Braun , B. K\"{o}rs and D. L\"{u}st, JHEP 0207 (2002) 026, hep-th/0206038 , hep-th/0210083;
G. Aldazabal, L. E. Ib\'{a}\~{n}ez, A.M. Uranga, hep-ph/0205250;
G. Honecker, JHEP 0201 (2002) 025, hep-th/0201037, hep-th/0112174;
H. Kataoka and M. Shimojo, hep-th/0112247; 
C. Kokorelis, hep-th/0207234,  hep-th/0205147;
J. R. Ellis, P. Kanti, D.V. Nanopoulos, hep-th/0206087 
G. Pradisi, hep-th/0210088

\bibitem{IRU}L. E. Ib\'{a}\~{n}ez, R. Rabad\'{a}n and A. M. Uranga, Nucl. Phys. B542 (1999) 112, hep-th/9808139

\bibitem{BKL2}D. Bailin, G. V. Kraniotis and A. Love, Phys. Lett. B547 (2002) 43, hep-th/0208103

\bibitem{BKL1}D. Bailin, G. V. Kraniotis and A. Love, Physics Letters B530 (2002) 202, hep-th/0108131







\bibitem{BKL4}D. Bailin, G. V. Kraniotis and A. Love, in preparation


\bibitem{cernyellow}Reports of the working groups on precision calculations for LEP2 Physics,
S. Jadach, G. Passarino, R. Pittau (Eds.), CERN Yellow Report, CERN, 2000-009

\bibitem{GIIQ} D. M. Ghilencia, L. E. Ib\'{a}\~{n}ez, N. Irges and F. Quevedo, hep-ph/0205083; 
E. Kiritsis and P. Anastasopoulos, JHEP 0205 (2002) 054, hep-ph/0201295; 
D. M. Ghilencia, hep-ph/0208205


\end{thebibliography}
\end{document}